\documentclass[letterpaper, 10 pt, conference]{ieeeconf}  

\IEEEoverridecommandlockouts                              
\usepackage{amsmath}
\usepackage{amssymb}  
\usepackage{amsfonts}
\usepackage[utf8]{inputenc}
\usepackage{tikz}
\usepackage{standalone}
\usepackage{xcolor}
\let\labelindent\relax
\usepackage[inline,shortlabels]{enumitem}
\usepackage{systeme}
\usepackage{amsthm}
\theoremstyle{assumption}
\newtheorem{assumption}{Assumption}
\theoremstyle{definition}
\newtheorem{definition}{Definition}

\usepackage{cancel}
\usepackage{tikz}
\usepackage{tcolorbox} 

\usepackage{hyperref}
\usepackage{cleveref}
\usepackage{siunitx}
\DeclareSIUnit[]{\vehicles}{veh}
\usepackage{pgfplots}
\usepackage{pgf}

\title{\LARGE \bf
{Control of a Mixed Autonomy Signalised Urban Intersection: \\ An Action-Delayed Reinforcement Learning Approach}
}

\author{Erica Salvato,  Arnob Ghosh, Gianfranco Fenu, and Thomas Parisini 
\thanks{E. Salvato and G. Fenu are with the Department of Engineering and Architecture of the University of Trieste
       {\tt\small erica.salvato@phd.units.it} and {\tt\small fenu@units.it}}
\thanks{A. Ghosh and T. Parisini are with the Dept. of Electrical and Electronic Engineering Department at Imperial College of London  
        {\tt\small arnob.ghosh@imperial.ac.uk} and {\tt\small t.parisini@imperial.ac.uk}}%
\thanks{
T. Parisini is also with the
Dept. of Engineering and Architecture, University of Trieste, 34127 Trieste, Italy, and  with the KIOS Research and Innovation
Center of Excellence, University of Cyprus, CY-1678 Nicosia, Cyprus.
}
\thanks{This work has been partially supported by European Union's Horizon 2020 research and innovation program under grant agreement no. 739551 (KIOS CoE) and by the Italian Ministry for Research in the framework of the 2017 Program for Research Projects of National Interest (PRIN), Grant no. 2017YKXYXJ.
}
}

\begin{document}

\maketitle

\begin{abstract}
We consider a mixed autonomy scenario where the traffic intersection controller decides whether the traffic light will be green or red at each lane for multiple traffic-light blocks. The objective of the traffic intersection controller is to minimize the queue length at each lane and maximize the outflow of vehicles over each block. We consider that the traffic intersection controller informs the autonomous vehicle (AV) whether the traffic light will be green or red for the future traffic-light block. Thus, the AV can adapt its dynamics by solving an optimal control problem. We model the decision process of the traffic intersection controller as a deterministic delay Markov decision process owing to the delayed action by the traffic controller. We propose Reinforcement-learning based algorithm to obtain the optimal policy. We show - empirically - that our algorithm converges and reduces the energy costs of AVs drastically as the traffic controller communicates with the AVs. 
\end{abstract}

\section{Introduction}


The autonomous vehicles (AVs) have the potential to disrupt the traffic-intersection control technology.  When there are only AVs, the traffic-intersection controller can plan the motions of the AVs in order to avoid collision at the intersection and provide the information to the AVs which then follow the dynamics. However, in the near future, the human-driven vehicles (HDVs) and AVs will co-exist. The traffic intersection controller can not communicate with the HDVs.  Rather the traffic intersection controller needs to rely on the traffic-lights to avoid collisions at the intersection. 

When there is a {\em mixed autonomy}, the traffic-intersection controller can not plan ahead since the behaviour of the HDVs are uncertain and may not follow any trajectory computed by the traffic-intersection controller. Hence, the question grasps the research community whether we can exploit the advantage of AVs in designing the traffic-intersection controller for {\em mixed autonomy}. The  advent of  artificial intelligence has the potential to revolutionise the traffic-light control algorithms as they can learn and adapt to the state of the traffic.  The pertinent question is can we develop a learning-based algorithm to control traffic-light duration  for mixed autonomy by exploiting the advantage of the AVs. 



In this paper, we consider a scenario where the traffic intersection controller decides whether the traffic-light will be green or red at each lane for each  traffic-light block (TLB) across a multiple time horizon(Section~\ref{sec:ps}). The objective is to minimise the queue-length at each lane and maximise the outflow of vehicles. Since the traffic intersection controller can communicate with the AV, we consider that the traffic intersection controller decides $d_a$ steps ahead from a certain time. Hence, the decision taken by the traffic intersection controller at the current time will be implemented at $d_a$ TLB ahead in the future. The traffic intersection controller communicates with the AVs of its decision. If an AV is not following any other vehicle or is far away from the preceding vehicle at a lane, it adapts its dynamics (Section~\ref{sec:av}) based on the decision. For example, if the traffic intersection controller informs that  the traffic-light will be green after $d_a$-time instances,  the AV adapts its dynamics to enter the intersection at the maximum speed at that instance, otherwise, the AV adapts its dynamics to stop at a distance from the intersection such that when the traffic light controller informs the time when the traffic-light will be green the AV can accelerate and enter the intersection at the maximum speed.  

We model the  decision process of the traffic intersection controller as a deterministic delayed Markov Decision Process (DDMDP) (Section~\ref{sec:rl}) since the traffic intersection controller's decision at a time instance is employed after a delay. The state is considered to be the queue length of the lanes. The reward is considered to be the difference between the queue lengths at two consecutive time instances. Computing an optimal decision is computationally challenging. The decision affects the dynamics of the vehicles in a non-linear non-smooth manner (Section~\ref{sec:IDM}) and one needs to compute the profile of the vehicles at every instance.  The behaviour of the HDVs, and the arrivals of the vehicles are difficult to predict.  Further, because of the delay in implementing the action, the traffic intersection controller is unaware of the exact state when the action will be implemented. We use a Reinforcement-learning (RL) based algorithm to learn the optimal policy for the traffic intersection controller using the $Q$-learning approach. Using the intelligent driver model (IDM), we simulate the vehicles' dynamics when the vehicles are HDV and when the AV is close to its preceding vehicle. When the AV is far from its preceding vehicle we propose an optimal control approach to compute the acceleration of the AV which depends on the decision of the traffic intersection controller by minimising the fuel-cost or the square integral of the change in velocity (Section~\ref{sec:av}).

We numerically investigate the proposed approach in two different scenarios (Section~\ref{sec:implementation}): i) AVs and HDVs coexist with equal proportion, and ii) only HDVs exist.
For both the scenarios, the algorithm converges in a reasonable time. The queue-lengths  are stabilised. Since the traffic intersection controller informs the leading AVs about the time where they can enter the intersection, our numerical analysis shows that total change in velocity of the AVs are very small compared to the corresponding values of the HDVs as the traffic intersection controller communicates with the AV. 





Algorithms to control the traffic-light duration at urban intersection  based on dynamic programming or  by control theory have been proposed, for instance, in~\cite{le_decentralized_2015,tettamanti_robust_2014,fleck_adaptive_2016,chiou_robust_2018,nilsson_micro-simulation_2020,liu_switching-based_2020}. See also~\cite{papageorgiou_review_2003,eom_traffic_2020} for broad reviews. Solutions using RL to cope with complex optimisation problems and uncertainties have been proposed in~\cite{hua,wiering2000multi,6083114,spall_traffic-responsive_1997,srinivasan_neural_2006,arel_reinforcement_2010,l_a_reinforcement_2011,chu_traffic_2017}. See also ~\cite{araghi_review_2015,eom_traffic_2020} for   broad reviews and~\cite{yau_survey_2017} for a recent survey about the use of RL techniques. However, all the above work did not consider the scenario where AVs can coexist with the HDVs. Recently, \cite{vinitsky2020optimizing, vinitsky2018benchmarks} consider RL-based algorithm for mixed autonomy.  However, all the above papers did not consider the possibility that the AVs can be informed about the time when they can enter the intersection. Compared to the above work, in our work, we model the decision process as a DDMDP. Further, we consider an optimal control algorithm for the AVs where they adapt to the information provided by the traffic intersection controller.  

Some authors  considered traffic-light-free intersection control designs when there are only AVs. See, e.g.,~\cite{zhang2016optimal,zhang2018penetration,malikopoulos2018decentralized} and the references therein. Several other authors proposed decentralised algorithm based on the coordination among the AVs  \cite{dresner2004multiagent,huang2012assessing}. However, in the near future AVs and HDVs will coexist, hence, those approaches can not be applied when there are HDVs since the traffic-light will be control the movement of the HDVs. Hence, we require new control mechanism for mixed autonomy. 
\section{System Model}
 \label{sec:ps}
We, first, describe the urban intersection system that we  consider (Section~\ref{sec:urban}). Subsequently, we describe the dynamics of the HDVs and AVs on the basis of the decision of the traffic intersection controller (Sections~\ref{sec:IDM} and \ref{sec:av}) which we use to compute the optimal decision of the traffic intersection controller in the next section (Section~\ref{sec:rl}). 


\subsection{The Urban Intersection system}\label{sec:urban}
\label{sec:urb_int}
\begin{figure}
    \centering
    \includegraphics[width=\linewidth]{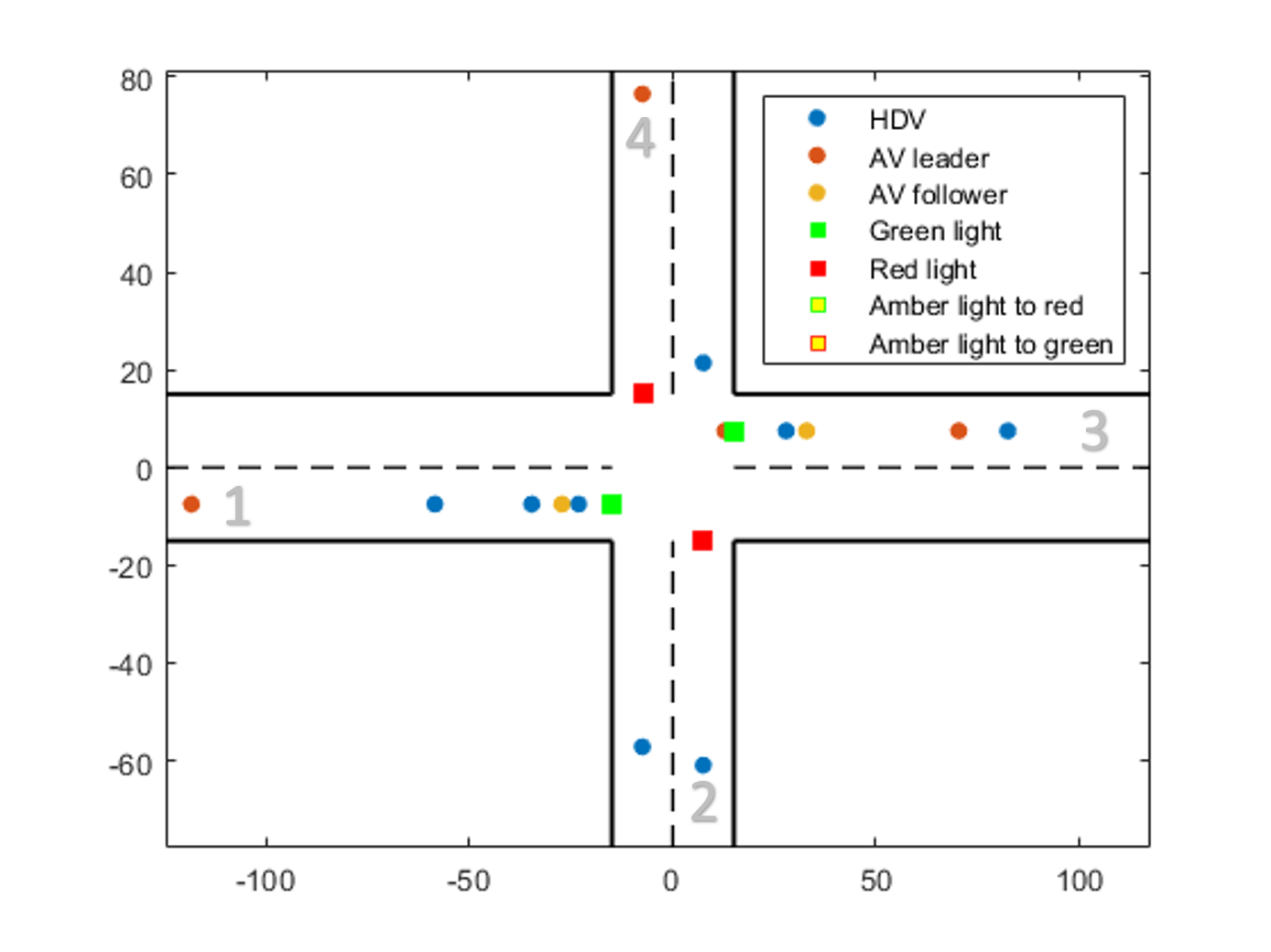}
    \caption{Traffic light controlled intersection in a mixed autonomy scenario}
    \label{fig:intersection}
\end{figure}

We consider a signalised urban intersection consisting of $4$ lanes (\Cref{fig:intersection}). 
We partition the urban intersection in \Cref{fig:intersection} in three main parts:
\begin{enumerate}
\item a \emph{Merging Zone} (MZ) of size $L_M \times L_M$, delimiting the area where vehicles of different lanes converge;
\item a \emph{Control Zone} (CZ) of length $L_C$ for each lane, where vehicles travel before entering the MZ;
\item an \emph{Exiting Zone} (EZ) of length $L_E$ for each lane, where vehicles travel after crossing the MZ. 
\end{enumerate}

A vehicle is considered to exit the intersection when it covers a distance greater than $L_C+L_M+L_E$. A traffic light is placed at the junction between the CZ and the MZ of each lane (4 traffic lights in total). Each vehicle enters the MZ when the respective traffic light is in a green status.
Conversely, a vehicle stops within the CZ when the respective traffic light exhibits a red status. Once the vehicles enter the MZ they cross the intersection. 

Each TLB is of fixed duration (e.g., few seconds to minutes). The traffic intersection controller decides whether the traffic-light will be green or red at a lane for each  TLB. We assume that the duration of a TLB is $T_{\text{RL}}$.  An amber light of fixed duration is added at each stage of traffic-light switching. Note that for consecutive TLBs, the traffic-light can be red or green at a lane depending on the decision of the traffic intersection controller. Thus, if for $m$-consecutive TLBs the traffic-light is red, then the traffic-light is red for $mT_{\text{RL}}$ duration.  

At $k$-th TLB, i.e., at $t_k$-th time the traffic intersection controller decides traffic-light for $k+d_a$-th TLB which will be implemented at $t_k+T_{\text{delay}}$. Hence $T_{\text{delay}}=d_aT_{\text{RL}}$  The delay is due to two reasons--i) the traffic intersection controller needs time to compute the optimal decision. We employ a Reinforcement Learning (RL) based algorithm to learn the optimal policy (Section~\ref{sec:rl}); ii) the traffic intersection controller provides AVs (if they are leading vehicle) the exact information at which TLB they can enter the intersection so that they can optimise their dynamics. The delay $T_{\text{delay}}$ ensures the AVs can adapt the dynamics. 
The current traffic signal is therefore the result of a past control input of the intersection controller. Note that the HDVs only follow the traffic-lights.

The goal of the RL-based traffic light controller to fulfil the following:
\begin{enumerate}
    \item minimise the queue length in each lane,
    \item maximise the rate of vehicles outflow.
\end{enumerate}

We, now, introduce some notations which we use throughout this section. 
We denote by $c(i,j)$ the $i$-th vehicle at lane $j$, where $i \in \{1,2,\hdots,N_\text{max}\}$, with $N_\text{max}$ maximum number of vehicles admitted in the intersection at the $j$-th lane, and $j \in \{1,2,\hdots,n_l\}$, with $n_l=4$ for the urban intersection in \Cref{fig:intersection}.

$v_{free}$ is the maximum allowable speed within the intersection system. A vehicle $c(i,j)$, entering the control zone of the $j$-th lane at time $t = t_{i,j}$ and travelling with a constant speed $v_{free}$ is supposed to entering the merging zone at:
\begin{equation}
    t_{i,j}^m=t_{i,j}+\frac{L_{C}}{v_{free}}.
\end{equation}
The above represents the time at which the vehicle $c(i,j)$ would enter the merging zone without any traffic. 

A vehicle $c(i,j)$, entering the control zone of the $j$-th lane at time $t = t_{i,j}$ and travelling with a constant speed $v_{free}$ is supposed to exit the merging zone at:
\begin{equation}
    t_{i,j}^f=t_{i,j}+\frac{L_C+L_M}{v_{free}}.
\end{equation}
The above represents the time at which the vehicle $c(i,j)$ would exit the merging zone without any traffic.  The above expressions are used to compute the wait-time of a vehicle which is the additional time the vehicle $c(i,j)$ spends at the intersection syste,.

We denote by $C^{(t_k)}$ the set of $N$ vehicles in all the lanes of the intersection system at $t = t_k$.
The HDVs set and the AVs set, respectively $C_{HD}^{(t_k)}$ and $C_{A}^{(t_k)}$, are both subsets of $C^{(t_k)}$: $C_{A}^{(t_k)}, C_{HD}^{(t_k)} \subseteq C^{(t_k)}$ s.t. $C_{A}^{(t_k)} \cap C_{HD}^{(t_k)} = \emptyset \land C_{A}^{(t_k)} \cup C_{HD}^{(t_k)} = C^{(t_k)}$.
Each vehicle maintains its own affiliation subset ($C_{HD}$ or $C_{A}$) for the entire cruise.

We denote with $p_{i,j}(t_k)$, $v_{i,j}(t_k)$, and $u_{i,j}(t_k)$ respectively the position, the speed, and the acceleration of the $c(i,j)$-th vehicle in the intersection at $t = t_k$. Each $i$-th vehicle entering the control zone of the $j$-th lane at $t=t_i$ will be initialised with an initial position $p_{i,j}(t_i)=0$.

\begin{definition}
\label{def:front}
Given two vehicles of the same $j$-th lane, $c(k,j)$ and $c(i,j)$, such that $p_{k,j}(t) > p_{i,j}(t)\; \forall t$, $k<i$ and $i>1$, then $c(i,j)$ is behind $c(k,j)$. When $k=i-1$ $c(k,j)$ is the \emph{front} vehicle of $c(i,j)$, i.e., the immediately preceding vehicle of $c(i,j)$.
\end{definition}

 

\begin{definition}
\label{def:non_conf}
A pair of lanes $(j,k)$  are non-conflicting if there are no intersection points that can lead to vehicles crashes. Let $\mathcal{L}$ be the set of non-conflicting pairs. 
\end{definition}

For example, the lane $1$ and $3$ are non-conflicting. However, the lanes $2$ and $1$ are conflicting. $\mathcal{L}=\{(1,3), (2,4)\}$. A traffic-controller can only make traffic-lights green simultaneously for the non-conflicting pair of lanes. 

Throughout the paper, the following assumptions take place.

\begin{assumption}
A vehicle $c(i,j) \in C$ can only go forward or stay still; i.e., no turning, backward gears, or lane changing are allowed.
\end{assumption}

\begin{assumption}
A vehicle $c(i,j) \in C_{A}$ is considered sensors equipped.
 $c(i,j)$ is able to estimate $p_{i-1,j}(t)$ and $v_{i-1,j}(t)$ if $c(i-1,j) \in C_{HD}$, while can access the actual values of $p_{i-1,j}(t)$ and $v_{i-1,j}(t)$ if $c(i-1,j) \in C_{A}$ when $(i-1,j)$ is the {\em front} vehicle (cf. Definition~\ref{def:front}) of $(i,j)$.
\end{assumption}
The first assumption can be relaxed by considering more complicated decision process and the dynamics. The second assumption entails that a AV can adapt to the preceding vehicle.



%


\subsection{Dynamics of Human Driven Vehicle}
\label{sec:IDM}
The dynamics of a HDV is described with the Intelligent Driver Model (IDM) \cite{treiber2000}.
It is an easy-to-tune adaptive cruise control system able to avoid vehicles collision in car-following mode.
The dynamics for a general $c(i,j)\in C_{HD}$ having a front vehicle $c(i-1,j)\in C$ is defined by:
\begin{equation}
\label{eq:IDM_follower_green}
\dot{v}_{i,j}(t)= u_{\text{max}}\left( 1-\left(\frac{v_{i,j}(t)}{\bar{v}_{i,j}}\right)^4-\frac{\left(s^*_{\{i,i-1\}}(t)\right)^2}{s^2_{\{i,i-1\}}(t)+\epsilon^2}\right),
\end{equation}
where $\bar{v}_{i,j}$ is the desired speed of the $(i,j)$-th vehicle, $u_{\text{max}}$ is the maximum acceleration, $s_{\{i,i-1\}}(t)= p_{i-1,j}(t)-p_{i,j}(t)$ is the current inter-vehicle distance, and $s^*_{\{i,i-1\}}$, the desired inter-vehicle distance:
\begin{eqnarray}
 s^*(v_{i,j}(t),\Delta v_{\{i,i-1\}}(t)) = s_{i,j}^0 + T_{i,j}v_{i,j}(t) + \nonumber\\
 \frac{v_{i,j}(t)\Delta v_{\{i,i-1\}}(t)}{2\sqrt{u_{\text{max}}u_{\text{min}}}},
\label{eq:dec_green}
\end{eqnarray}
where $\Delta v_{\{i,l\}}(t)=v_{l,j}(t)-v_{i,j}(t)$ is the vehicles difference in speed, $u_{\text{min}}$ is the maximum deceleration, $s_{i,j}^0$ is the jam distance and $T_{i,j}$ the safety time gap between two vehicles. 



The dynamics of the vehicle $c(i,j)$  when it  observes a red-light and  the vehicle $c(i-1,j)$ passes the CZ, then the  vehicle $c(i,j)$  then needs to decelerate irrespective of the dynamics of the vehicle $c(i-1,j)$. Here, we consider the red-light as a static vehicle situated at the end of the CZ of the lane. Thus, $\Delta v_{\{i,TL\}}(t)=-v_{i,j}(t)$, while $s_{\{i,TL\}}(t)=p_{TL}(t)-p_{i,j}(t)$ is the current distance of the vehicle $c(i,j)$ from the traffic light where $p_{TL}(t)=L_C$ and $v_{TL}(t)=0$ for all $t$. Thus, the last term in the right-hand side of \Cref{eq:IDM_follower_green}  is replaced by the following term
\begin{equation} 
 \frac{\left(s^*_{\{i,TL\}}\right)^2}{s^2_{\{i,TL\}}(t)+\epsilon^2}
\label{eq:dec_red}
\end{equation}
where $s^*_{\{i,TL\}}$ is the desired distance from the traffic light with $\Delta v_{\{i,TL\}}(t)=-v_{i,j}(t)$ in place of $\Delta v_{\{i,i-1\}}(t)$. 

When the vehicle $c(i,j)$ is   $d_{follow}$ distance away from the preceding vehicle or there is no other vehicle in the lane, then the reaction of the vehicle $c(i,j)$ would not be sensitive to the dynamics of the preceding vehicle. An HDV $c(i,j)$ having an inter-vehicle distance $s_{\{i,l\}}(t)>d_{\text{follow}}$ and observing a green light follows \Cref{eq:IDM_follower_green} with $s_{\{i,i-1\}}(t)=\infty$. Thus, it would behave like the preceding vehicle situated at the infinity. On the other hand the vehicle $c(i,j)$ facing the red-light will follow \Cref{eq:IDM_follower_green} with $s_{\{i,i-1\}}(t)=\infty$ till the distance from the MZ becomes less than $d_{\text{follow}}$. Since the red-light is considered to be a static vehicle at the position $p_{TL}(t)=L_c$, hence the dynamics of the vehicle $c(i,j)$ will be the same as the \Cref{eq:IDM_follower_green} where the last term is replaced by (\ref{eq:dec_red}) as described in the previous paragraph when $p_{TL}(t)-p_{i,j}(t)\leq d_{\text{follow}}$ and the traffic-light is red. 

We are only left to describe the dynamics when the vehicle $c(i,j)$ is the lead vehicle in the CZ and faces the amber light. If $p_{i,j}(t)-p_{TL}(t)\leq d_{\text{follow}}$, then the vehicle $c(i,j)$ will follow \Cref{eq:IDM_follower_green} with the preceding vehicle either at $\infty$ or at $p_{i,j}(t)$ if $p_{i-1,j}(t)-p_{i,j}(t)\leq d_{\text{follow}}$. If $p_{i,j}(t)-p_{TL}(t)> d_{\text{follow}}$, then the vehicle $c(i,j)$ will follow  the scenario where the traffic-light is red. Intuitively, if the amber-light is switched on and the vehicle $c(i,j)$ is very close to the intersection, it will follow its dynamics, otherwise, the vehicle would decelerate to stop. 


\subsection{Autonomous vehicle}\label{sec:av}

The AV $c(i,j)$ enters the CZ at time $t_{i,j}^0$. If an AV $c(i,j)$ is far away from its preceding vehicle $c(i-1,j)$, i.e., $p_{i-1,j}(t)-p_{i,j}(t)>d_{\text{follow}}$ or if it is the lead vehicle then the AV moves at the constant speed, till the time reaches $t_k$,i.e., $t_k=T_{\text{RL}}\lceil t/T_{\text{RL}}\rceil$, when  the the traffic-intersection controller decides whether the traffic-light will be green or red at the $k+d_a$-th TLB at each lane. The traffic-controller then broadcasts the information to all the AVs. Since each TLB is of $T_{\text{RL}}$ duration, thus, the decision is implemented at time $t_k+d_aT_{\text{RL}}$. Recall that $T_{\text{delay}}=d_aT_{\text{RL}}$. The AV then computes its acceleration/deceleration solving an optimisation problem which we describe next. 

Specifically, the AV would want to find the control input acceleration/deceleration $u_{i,j}(t)$ which governs the dynamics as follows
\begin{equation}
\label{eqn:AVlead}
\begin{aligned}
\dot{v}_{i,j}(t) &= u_{i,j}(t) \\
\dot{p}_{i,j}(t) &= v_{i,j}(t),
\end{aligned}
\end{equation}
We now describe how the AV $c(i,j)$ computes $U_{i,j}(t)$
When the traffic-light at any lane changes from the $k+d_a-1$-th period to $k+d_a$-th period then the amber light of duration $T_{\text{alert}}$ is applied. 

If at $t = t_k$ the AV with $p_{i,j}(t_k)=p_{t_k}$ and  $v_{i,j}(t_k)=v_{t_k}$, is informed by the intersection controller that at $t = t_k+T_{\text{delay}}$ the traffic light will be set to a green status, then $u_{i,j}(t)$ will be the solution of the following optimisation:
\begin{equation}
\begin{aligned}
  \mathcal{P}_1:  \min_{u_{i,j}(t)} & \;\frac{1}{2}\int_{t_k}^{t_\text{cross}} u_i^2(t) dt,\\
    \text{subject to:} & \;(\ref{eqn:AVlead}),\; p_{i,j}(t_k)=p_{t_k},\; v_{i,j}(t_k)=v_{t_k},\\ & v_{i,j}(t)\leq v_{\text{max}}\;\forall\;t\;\in\; \left[t_k,t_\text{cross}\right],\\ 
    & v_{i,j}(t)\geq 0\;\forall\;t\;\in\; \left[t_k,t_\text{cross}\right],\\ & p_{i,j}(t_\text{cross})=L_C, \\& t_\text{cross} \geq t_k+T_{\text{delay}}+T_{\text{alert}}\\
    & t_\text{cross} \leq t_k+T_{\text{delay}}+T_{RL}.
\end{aligned}
\label{eq:cost1}
\end{equation}
The objective indicates that the AV tries to minimise the total energy cost which is represented as the integral of the square of $u_{i,j}(t)$. The first constraint represents the initial conditions and the vehicle dynamics. $t_{\text{cross}}$ is the time at which the AV will enter the MZ, hence, the velocity must follow the constraints that the lower bound of the velocity is $0$ and the upper bound of velocity is $v_{max}$. Since $t_{\text{cross}}$ is the time at which the vehicle must enter the MZ, thus, the position at that time must be $L_C$. The penultimate equation represents that the vehicle can not enter the MZ before the start of the TLB $k+d_a$, hence $t_\text{cross} \geq t_k+T_{\text{delay}}+T_{\text{alert}}$. Note that $T_{\text{alert}}$ is $0$, if the decision of the traffic intersection controller is the same for the lane $j$ for $k-1$-th and $k$-th periods. The final constraint ensures that the vehicle must cross the intersection before the end of the $k+d_a$-th period, i.e., before the start of the $k+d_a+1$-th period. Note that the decision variable is $u_{i,j}(t)$ and $t_{\text{cross}}$. 

In order to obtain an optimal solution, we relax the constraints and add the following penalties in the objective
\begin{align}
    K_{v_{\text{max}}}& \{ \max(0,(v_{i,j}(t)-v_{\text{max}}))\},\\
    K_{v_{\text{min}}}& \{ \max(0,-v_{i,j}(t))\},\\
    K^1_{t_{\text{cross}}}& \{ \max(0,(t_k+T_{\text{delay}}+T_{\text{alert}}) - t_{\text{cross}})\},\\
    K^2_{t_{\text{cross}}}& \{ \max(0,(t_\text{cross}- (t_k+T_{\text{delay}}+T_{\text{alert}}))\}.
\end{align}
to the cost in \Cref{eq:cost1}. For $t\geq t_{\text{cross}}$ we set a constant maximum speed profile ($v_{i,j}(t)=v_{\text{max}}$), thus allowing an increase of the throughput. The delay $d_a$ is chosen such that there always exists a solution of the optimisation problem. Note that if we increase $d_a$, we are guaranteed to obtain an optimal solution, however, high value of $d_a$ may not serve its purpose. 

In contrast, if at $t = t_k$ the AV $(i,j)$, having  $p_{i,j}(t_k)=p_{t_k}$ and a $v_{i,j}(t_k)=v_{t_k}$, is informed by the intersection controller that at $ t_{\text{stop}}=t_k+T_{\text{delay}}$ the traffic light will be set to a red status, then the deceleration will be the solution of the following optimisation:
\begin{equation}
\begin{aligned}
  \mathcal{P}_2:   \min_{u_i(t)} & \;\frac{1}{2}\int_{t_k}^{t_\text{stop}} u_{i,j}^2(t) dt,\\
    \text{subject to:} & \;(\ref{eqn:AVlead}),\; p_{i,j}(t_k)=p_{t_k},\; v_{i,j}(t_k)=v_{t_k},\\ & p_{i,j}(t_\text{stop})=L_C-\delta_a,  v_{i,j}(t_\text{stop}) = 0,\\& v_{i,j}(t)\leq v_{\text{max}}\;\forall\;t\;\in\; \left[t_k,t_\text{stop}\right],\\ 
    & v_{i,j}(t)\geq 0\;\forall\;t\;\in\; \left[t_k,t_\text{stop}\right].
\end{aligned}
\label{eq:cost2}
\end{equation} 
Also in this case we relax the optimisation problem by relying on a penalty function approach. 

Compared to $\mathcal{P}_1$, in $\mathcal{P}_2$, the AV $(i,j)$ stops at time $t_{\text{stop}}$. Further, the AV stops at $\delta_a$ distance from the MZ. It will ensure that when the traffic-light becomes green, the AV can accelerate and enter the intersection at the maximum speed. The optimal solution always exists and is computed numerically via MATLAB. 

The AV $c(i,j)$ will follow the dynamics either computed by relaxed version of $\mathcal{P}_1$ or by relaxed version of $\mathcal{P}_2$ depending on the decision of the traffic-intersection controller. At $k^{\prime}>k$-th period, the vehicle $c(i,j)$ updates its dynamics if the traffic intersection controller decides to make the traffic-light as green at the $k^{\prime}$-th period (which will be implemented at the $k^{\prime}+d_a$-th TLB) if the decision of the traffic intersection controller at the $k$-th period was to make the traffic light at red at lane $j$ for $k+d_a$-th TLB. In this case, the vehicle $c(i,j)$ again solves the optimisation problem $\mathcal{P}_1$ starting from $t_{k^{\prime}}$ with $t_{k^{\prime}}$ replacing $t_k$ for each of the constraints of $\mathcal{P}_1$. The vehicle $c(i,j)$ will enter the intersection inside the $k^{\prime}+d_a$-th TLB. 

Once the vehicle $c(i,j)$ gets the information that the traffic-light will be green from the traffic intersection it will not update its dynamics unless $p_{i-1,j}(t)-p_{i,j}(t)\leq d_{\text{follow}}$ at a certain time $t$, i.e., the vehicle $c(i,j)$ gets close to its preceding vehicle. In this case, the AV will discard the solution returned by the optimisation problem and the information from the traffic intersection controller and computes the acceleration/deceleration according to the IDM model as described in Section~\ref{sec:IDM} from that time onwards throughout the journey. 




\section{Decision Making strategy}\label{sec:rl}
In this section, we describe how the intersection controller takes its decision using a RL-based algorithm.
Henceforth, we assume that vehicles' dynamics are discretised with a sampling time $T_S$, while the intersection controller works at fixed time-steps $T_{\text{RL}}=nT_S$, with $n \in \mathbb{N}$ fixed. 
Therefore, the amber traffic light duration and the RL agent delay, already mentioned in \Cref{sec:urb_int}, are respectively $T_{\text{alert}}=mT_S$, with $m<n \in \mathbb{N}$ fixed. $T_{\text{RL}}$ is exactly equal to the length  the duration of TBL. 

We characterise the decision process for urban intersection traffic-light control as a DDMP \cite{Katsikopoulos2003} with constant action and cost delays.

The considered DDMDP is a tuple $\langle S,A,p,g,d_a,d_c \rangle$ where $S$ is the state set, $A$ is the control input (\emph{action}) set, $p$ the transition probability describing the dynamics of the system to be controlled, $g:\;S\times A\times S\rightarrow \mathbb{R}$ the reward function quantifying the benefit of a certain action choice given a particular state of the process (\emph{reward}), $d_a$ the constant action delay, and $d_c=d_a$ the constant delay to observe the reward.

The controller at each $k$-th time instance observes the state of the system and has to select a control input that will be applied at the $(k+d_a)$-th step. 
This means that, at the $k$-th time instance the action selected at the $(k-d_a)$-th time instant is applied. The reward  is measured after $k+d_a+1$-th time instance when the action is decided at the $k$-th time instance.  
As shown in \cite{bertsekas1995dynamic} a DDMDP with action and cost delay is reducible to a classical MDP without dealy.
The equivalent MDP is a tuple $\langle I,A,p,g_a \rangle$,
where $I:S\times A^{d_a}$ and $g_a:S\times A^{d_a}\times A\times S\times A^{d_a}\rightarrow \mathbb{R}$ are respectively the "modified" state set and the ``modified" reward function. 

Throughout the paper we denote:
\begin{itemize}
    \item $s^{(k)} \in S$ the state of the RL system at the $k$-th time instant;
    \item $a^{(k)} \in A$ the RL control input (\emph{action}) at the $k$-time instant;
    \item $I^{(k)}=(s^{(k)},a^{(k-d_a)},\dots,a^{(k-1)}) \in I$ the information needed for optimal action selection at the $k$-th time instant;
    \item $g_a(i^{(k)},a^{(k)}, i^{(k+1)})=g(s^{(k)},a^{(k-d_a)},s^{(k+1)})$ the reward function.
\end{itemize} 
Note that the reward function at the $k$-th instance does not depend on the action at the $k$-th instance since the reward at the $k$-th instance depends on the action decided at the $k-d_a$-th instance $a^{(k-d_a)}$.  The transition probability $p$ for the original DDMDP can be extended to this modified DDMP by observing the following  $p(I^{(k+1)}|I^{(k)},a)=p(s^{(k+1)}|s^{(k)},a)$ for all $e_1^TI^{(k+1)}=s^{(k+1)}$, $e_1^TI^{(k)}=s^{(k)}$, and $e_{k+1}^TI^{(k)}=a$, the rest of elements are $0$, where $e_{m}$ is the unit vector with only the $m$-th component is $1$, and the rest are $0$.  

\subsection{State Action and Reward in the urban intersection}
We now characterise the state, the action and the reward in our setting. First, we introduce some notations. 

Hereinafter we will refer to $\mathbb{N} \cup \{0\}$ as $\mathbb{N}_+$; and to $\mathbb{B}=\{0;1\}$ as the Boolean domain. 


\subsubsection{State}
The state $s^{(k)} \in S$ at the $k$-th time instant is equal to $X^{(k)}$,
where:
 $X^{(k)} \in {\mathbb{N}_+}^{n_l}$ is a vector in which the $j$-th element $x^{(k)}_{j}$ element represents the number of vehicles in the CZ of the $j$-th lane.

\subsubsection{Action}
The action $a^{(k)} \in A$ at the $k$-th time instant is a vector having the number of elements equal to the number of lanes $A:=\mathbb{B}^{n_l}$.
At the $k$-th instance, the traffic controller decides which lane to be open, i.e., for which lane the traffic-light will be green at the $k+d_a$-th TLB. If $a_j^{(k)}=1$, the traffic-light will be green, if $a_j^{(k)}=0$, the traffic-light will be red for lane $j$ at the $k+d_a$-th TLB. Note that only those lanes which are non-conflicting can be open simultaneously.
Therefore, we reduce the problem imposing that the actions of non-conflicting lanes are equal.Hence, $a_j^{(k)}=a_{l}^{(k)}$
if the pair $(j,l)$ are non-conflicting. Thus, the action space can be reduced to only choosing elements for the set $\mathcal{L}$, i.e., the non-conflicting lanes. 

\subsubsection{Reward function}
The traffic-intersection controller wants to minimise the queue length at each lane, and maximise the outflow of vehicles at a given instance. Hence, we consider the reward-function as the following
Thus, maximising 
\begin{equation}
    g(s^{(k)},a^{(k-d_a)},s^{(k+1)})=\parallel W X^{(k)} \parallel_1 - \parallel W X^{(k+1)} \parallel_1,
\end{equation}
where $W\in \mathbb{R}^{n_l}$ is the weight vector.  The weights vector $W$ allows to assign to each lane a relative priority. 
If we want to impose a higher priority for the $j$-th lane, we will assign to the $j$-th element of $W$ ($w_j$) a higher value than the others ($w_i<w_j,\;\forall i \neq j$).
Imposing $W=[1,1,\hdots,1]$ each lane assume the same relevance in the optimisation.

\subsection{Optimal Policy and Q-Learning}
The traffic-intersection controller has to learn an optimal control law (\emph{policy}) $\pi:I\rightarrow A$; i.e., the one that maximises the expected value of the discounted cumulative reward given the state $I^{(k)}$. We also assume that $d_a$ decided actions are already awaiting execution at $k=0$ which constitutes the initial state $I^{(0)}$. 

\begin{equation}
    \mathbb{E}\left[\sum\limits_{k=0}^{H}\gamma^k g(s^{(k)},a^{(k-d_a)},s^{(k+1)})\right],
\end{equation}
where $\gamma \in [0,1]$ is a discounted factor (a constant real value quantifying how much important the future reward is compared to the immediate one) and $H$ is the horizon of the optimisation problem to be solved.

In the following we rely on a tabular Q-Learning algorithm \cite{watkins1992q}: an off-policy value function approach that uses Q-values. Since we have only used a simulation model to simulate the vehicle dynamics, thus, off-policy evaluation is not costly.

Note that in order to find the optimal policy $\pi$, we need to evaluate the $Q$-function for the modified DDMP. Hence, we will compute $Q(I^{(k)},a^{(k)})$ for all $(I^{(k)}, a^{(k)})$ and then finding the optimal policy as
\begin{align}
    \pi^{*}(I^{(k)})=\arg\max_{a^{(k)}}Q(I^{(k)},a^{(k)}). 
\end{align}
The reward inherently depends on the dynamics of the vehicles which have been described in Sections~\ref{sec:IDM} and \ref{sec:av} which in turn depend on the decision of the traffic intersection controller. The dynamics is non-linear and discontinuous as well. 

Note that as $d_a$ increases the state-space will increase, however, it will help to obtain a solution for the AV's optimisation problem. 

\section{Implementation}\label{sec:implementation}
\label{sec:exp}
To evaluate the proposed approach we design a MATLAB framework.

The considered simulation example consists of the $4$ lanes intersection described in \Cref{sec:ps} ($n_l=4$) having a merging zone of size $L_M=\SI{30}{\meter}$, while the control zone length $L_C$ and the exit zone length $L_E$ of each lane are both equal to $\SI{400}{\meter}$. 
The maximum speed limit is set to $v_{\text{max}}=\SI{13}{\meter \per\second}$. The vehicle arrival rate follows a Poisson process with $\lambda=\SI{450}{\vehicles\per\hour}$. Each vehicle is set to enter the intersection with an initial position $p_{i,j}(t_0)=0$ (corresponding to the maximum distance from the merging zone), an initial speed $v_{i,j}(t_0)$ randomly sampled in $[\SI{9}{\meter \per\second},\SI{11}{\meter \per\second}]$, and an initial acceleration $u_{i,j}(t_0)$ randomly sampled in $[\SI{0}{\meter \per\second\squared },\SI{0.5}{\meter \per\second\squared }]$.
We set the capacity of the intersection equal to $N_{\text{max}}=100$, i.e., if there are $100$ vehicles at the intersection, the intersection will be closed for access from any lane.  We randomly set the amount of AVs entering the intersection in order to compare the proposed approach on two different scenarios:
\begin{enumerate}
    \item $50\%$ of AVs and $50\%$ of HDVs;
    \item only HDVs.
\end{enumerate}
We consider the same jam distance and the same safety time gap for each vehicle modelled as an IDM (respectively $s_{i,j}^0=\SI{2}{\meter}$ and $T_{i,j}=\SI{5}{\second}$), and we set the $\epsilon=\SI{1.6}{\meter}$ in order to avoid an high deceleration when the bumper-to-bumper distance of two subsequent vehicles is close to zero.
We set $\delta_a=\SI{12}{\meter}$.
It allows a soft acceleration profile once the AV receives the information that the traffic light will be green. We set $d_{\text{follow}}=\SI{5}{\meter}$. 

The RL controller training is performed solving an infinite horizon Q-Learning problem and starts with an intersection having zero vehicles. We set $W=[1,1,\hdots,1]$, hence, the controller did not prioritise among the lanes. 

\begin{figure}
    \centering
    \resizebox{\linewidth}{!}{
    \input{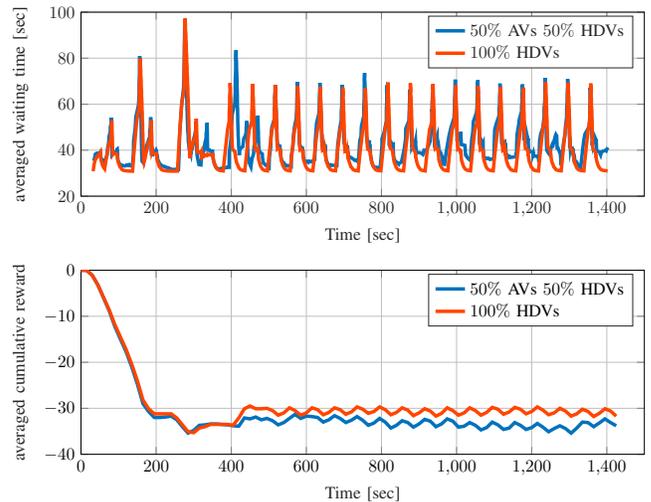}}
    \caption{Averaged vehicles waiting time in the intersection lanes during the simulations (top) and averaged cumulative reward (bottom) during the simulations.} 
    \label{fig:RewandWait}
\end{figure}
\begin{figure}
    \centering
    \resizebox{\linewidth}{!}{\input{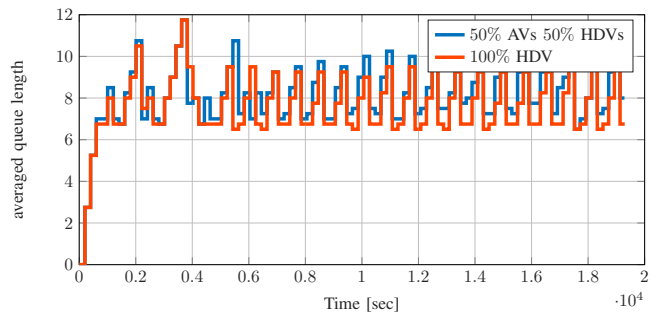}}
    \caption{Averaged queue length in the intersection lanes during the simulations.} 
    \label{fig:avQueue}
\end{figure}
We performed both the simulation considering $T_{\text{RL}}=\SI{15}{\second}$, $T_{\text{delay}}=2T_{\text{RL}}=\SI{30}{\second}$, and $T_{\text{alert}}=\SI{3}{\second}$, i.e, $d_a=2$.
In \Cref{fig:RewandWait}  the average vehicles' waiting time and the averaged cumulative reward of the performed trainings are shown respectively. In particular the blue plots are related to the scenario of an equally distributed number of AVs and HDVs in the intersection, while the orange depict the result obtained in the scenario of only HDVs. 
We can observe that in both the simulations the vehicles waiting time assumes a pseudo-periodic behaviour in correspondence of reward.
However, no significant differences between the two different scenarios can be observed as also highlighted in \Cref{fig:avQueue}, where the respective average lane queues are shown.
\begin{table*}[]
\centering
\caption{Statistics of the $\int u^2(t)$ for coexisting AVs and HDVs}
\begin{tabular}{|c|l|c|c|c|c|}
                                           \hline                    & \textbf{vehicle} & \textbf{mean} & \textbf{median} & \textbf{mode}
                                           & \begin{tabular}{c}\textbf{standard}\\ \textbf{deviation}\end{tabular} \\\hline

\begin{tabular}{c}\textbf{50\%} \textbf{AVs}\\\textbf{ 50\%} \textbf{HDVs}\end{tabular} & AVs  & 7.5483 & 0.9414 & 6.6331e-05 & 13.4863\\
                    & & 1.3509e+04 & 211.6577 & 2.8527 & 2.5678e+04\\\hline
\textbf{100\% HDV}s & HDVs & 1.6233e+03 & 85.5469 & 1.1249     & 4.4520e+03  \\\hline                            
\end{tabular}
\label{tab:u2_int_Comp}
\end{table*}


On the other hand, as expected, the clear advantage of introducing AVs can be observed by comparing the energy consumption of the two different vehicles. 
Indeed, as highlighted in \Cref{tab:u2_int_Comp}, where for both vehicle types some statistical indices of the integral of the squared acceleration profiles over time are shown, AVs outperform HDVs.
Moreover, comparing only the statistical HDVs values of this integral in both scenarios, in the $100\%$ HDVs setting, respectively, the mean and standard deviation take a higher value.
Therefore, we can conclude that the introduction of AVs also decreases the total energy of HDVs.

\section{Conclusion and Future Work}
We model the decision process of a traffic intersection controller as a DDMDP and propose a RL-based algorithm to compute the optimal policy to decide whether the traffic-light will be green or red at each lane over a long-time period. The decision taken by the traffic-light controller is applied after a delay which is communicated to the AVs. Thus, the AVs can adapt their dynamics in a better way. Numerically, we show that the AVs can in fact minimise total acceleration and deceleration compared to the HDV without increasing the wait-time due to the delayed decision process.

The characterisation of the optimal policy for different set of rewards are left for the future. The consideration of the scenario where the AVs have an unreliable communication with the traffic-intersection constitutes an important future research direction. The investigation of the delay parameter on the performace has been left for the future. 

\bibliographystyle{IEEEtran}
\bibliography{ref.bib}

\begin{thebibliography}{10}
\providecommand{\url}[1]{#1}
\csname url@samestyle\endcsname
\providecommand{\newblock}{\relax}
\providecommand{\bibinfo}[2]{#2}
\providecommand{\BIBentrySTDinterwordspacing}{\spaceskip=0pt\relax}
\providecommand{\BIBentryALTinterwordstretchfactor}{4}
\providecommand{\BIBentryALTinterwordspacing}{\spaceskip=\fontdimen2\font plus
\BIBentryALTinterwordstretchfactor\fontdimen3\font minus
  \fontdimen4\font\relax}
\providecommand{\BIBforeignlanguage}[2]{{%
\expandafter\ifx\csname l@#1\endcsname\relax
\typeout{** WARNING: IEEEtran.bst: No hyphenation pattern has been}%
\typeout{** loaded for the language `#1'. Using the pattern for}%
\typeout{** the default language instead.}%
\else
\language=\csname l@#1\endcsname
\fi
#2}}
\providecommand{\BIBdecl}{\relax}
\BIBdecl

\bibitem{le_decentralized_2015}
T.~Le, P.~Kov{\'a}cs, N.~Walton, H.~L. Vu, L.~L. Andrew, and S.~S. Hoogendoorn,
  ``Decentralized signal control for urban road networks,''
  \emph{Transportation Research Part C: Emerging Technologies}, vol.~58, pp.
  431--450, 2015.

\bibitem{tettamanti_robust_2014}
T.~Tettamanti, T.~Luspay, B.~Kulcsar, T.~Peni, and I.~Varga, ``Robust {Control}
  for {Urban} {Road} {Traffic} {Networks},'' \emph{IEEE Transactions on
  Intelligent Transportation Systems}, vol.~15, no.~1, pp. 385--398, 2014.

\bibitem{fleck_adaptive_2016}
J.~L. Fleck, C.~G. Cassandras, and Y.~Geng, ``Adaptive {Quasi}-{Dynamic}
  {Traffic} {Light} {Control},'' \emph{IEEE Transactions on Control Systems
  Technology}, vol.~24, no.~3, pp. 830--842, 2016.

\bibitem{chiou_robust_2018}
S.-W. Chiou, ``A robust signal control system for equilibrium flow under
  uncertain travel demand and traffic delay,'' \emph{Automatica}, vol.~96, pp.
  240--252, 2018.

\bibitem{nilsson_micro-simulation_2020}
G.~Nilsson and G.~Como, ``A {Micro}-{Simulation} {Study} of the {Generalized}
  {Proportional} {Allocation} {Traffic} {Signal} {Control},'' \emph{IEEE
  Transactions on Intelligent Transportation Systems}, vol.~21, no.~4, pp.
  1705--1715, 2020.

\bibitem{liu_switching-based_2020}
D.~Liu, W.~Yu, S.~Baldi, J.~Cao, and W.~Huang, ``A {Switching}-{Based}
  {Adaptive} {Dynamic} {Programming} {Method} to {Optimal} {Traffic}
  {Signaling},'' \emph{IEEE Transactions on Systems, Man, and Cybernetics:
  Systems}, vol.~50, no.~11, pp. 4160--4170, 2020.

\bibitem{papageorgiou_review_2003}
M.~Papageorgiou, C.~Kiakaki, V.~Dinopoulou, A.~Kotsialos, and {Yibing Wang},
  ``Review of road traffic control strategies,'' \emph{Proceedings of the
  IEEE}, vol.~91, no.~12, pp. 2043--2067, 2003.

\bibitem{eom_traffic_2020}
M.~Eom and B.-I. Kim, ``The traffic signal control problem for intersections: a
  review,'' \emph{European Transport Research Review}, vol.~12, no.~1, p.~50,
  2020.

\bibitem{hua}
\BIBentryALTinterwordspacing
H.~Wei, G.~Zheng, H.~Yao, and Z.~Li, ``Intellilight: A reinforcement learning
  approach for intelligent traffic light control,'' in \emph{Proceedings of the
  24th ACM SIGKDD International Conference on Knowledge Discovery \& Data
  Mining}, ser. KDD '18.\hskip 1em plus 0.5em minus 0.4em\relax New York, NY,
  USA: Association for Computing Machinery, 2018, p. 2496–2505. [Online].
  Available: \url{https://doi.org/10.1145/3219819.3220096}
\BIBentrySTDinterwordspacing

\bibitem{wiering2000multi}
M.~A. Wiering, ``Multi-agent reinforcement learning for traffic light
  control,'' in \emph{Machine Learning: Proceedings of the Seventeenth
  International Conference (ICML'2000)}, 2000, pp. 1151--1158.

\bibitem{6083114}
M.~Abdoos, N.~Mozayani, and A.~L.~C. Bazzan, ``Traffic light control in
  non-stationary environments based on multi agent q-learning,'' in \emph{2011
  14th International IEEE Conference on Intelligent Transportation Systems
  (ITSC)}, 2011, pp. 1580--1585.

\bibitem{spall_traffic-responsive_1997}
J.~C. Spall and D.~C. Chin, ``Traffic-responsive signal timing for system-wide
  traffic control,'' in \emph{1997 {American} {Control} {Conference} ({ACC})},
  vol.~4, 1997, pp. 2462--2463.

\bibitem{srinivasan_neural_2006}
D.~Srinivasan, M.~C. Choy, and R.~L. Cheu, ``Neural {Networks} for
  {Real}-{Time} {Traffic} {Signal} {Control},'' \emph{IEEE Transactions on
  Intelligent Transportation Systems}, vol.~7, no.~3, pp. 261--272, 2006.

\bibitem{arel_reinforcement_2010}
I.~Arel, C.~Liu, T.~Urbanik, and A.~G. Kohls, ``Reinforcement learning-based
  multi-agent system for network traffic signal control,'' \emph{IET
  Intelligent Transport Systems}, vol.~4, no.~2, pp. 128--135, 2010.

\bibitem{l_a_reinforcement_2011}
P.~L.~A. and S.~Bhatnagar, ``Reinforcement {Learning} {With} {Function}
  {Approximation} for {Traffic} {Signal} {Control},'' \emph{IEEE Transactions
  on Intelligent Transportation Systems}, vol.~12, no.~2, pp. 412--421, 2011.

\bibitem{chu_traffic_2017}
T.~Chu and J.~Wang, ``Traffic signal control by distributed {Reinforcement}
  {Learning} with min-sum communication,'' in \emph{2017 {American} {Control}
  {Conference} ({ACC})}, 2017, pp. 5095--5100.

\bibitem{araghi_review_2015}
S.~Araghi, A.~Khosravi, and D.~Creighton, ``A review on computational
  intelligence methods for controlling traffic signal timing,'' \emph{Expert
  Systems with Applications}, vol.~42, no.~3, pp. 1538--1550, 2015.

\bibitem{yau_survey_2017}
K.-L.~A. Yau, J.~Qadir, H.~L. Khoo, M.~H. Ling, and P.~Komisarczuk,
  ``\BIBforeignlanguage{en}{A {Survey} on {Reinforcement} {Learning} {Models}
  and {Algorithms} for {Traffic} {Signal} {Control}},''
  \emph{\BIBforeignlanguage{en}{ACM Computing Surveys}}, vol.~50, no.~3, pp.
  1--38, 2017.

\bibitem{vinitsky2020optimizing}
E.~Vinitsky, N.~Lichtle, K.~Parvate, and A.~Bayen, ``Optimizing mixed autonomy
  traffic flow with decentralized autonomous vehicles and multi-agent rl,''
  2020.

\bibitem{vinitsky2018benchmarks}
E.~Vinitsky, A.~Kreidieh, L.~Le~Flem, N.~Kheterpal, K.~Jang, C.~Wu, F.~Wu,
  R.~Liaw, E.~Liang, and A.~M. Bayen, ``Benchmarks for reinforcement learning
  in mixed-autonomy traffic,'' in \emph{Conference on Robot Learning}.\hskip
  1em plus 0.5em minus 0.4em\relax PMLR, 2018, pp. 399--409.

\bibitem{zhang2016optimal}
Y.~J. Zhang, A.~A. Malikopoulos, and C.~G. Cassandras, ``Optimal control and
  coordination of connected and automated vehicles at urban traffic
  intersections,'' in \emph{2016 American Control Conference (ACC)}.\hskip 1em
  plus 0.5em minus 0.4em\relax IEEE, 2016, pp. 6227--6232.

\bibitem{zhang2018penetration}
Y.~Zhang and C.~G. Cassandras, ``The penetration effect of connected automated
  vehicles in urban traffic: an energy impact study,'' in \emph{2018 IEEE
  Conference on Control Technology and Applications (CCTA)}.\hskip 1em plus
  0.5em minus 0.4em\relax IEEE, 2018, pp. 620--625.

\bibitem{malikopoulos2018decentralized}
A.~A. Malikopoulos, C.~G. Cassandras, and Y.~J. Zhang, ``A decentralized
  energy-optimal control framework for connected automated vehicles at
  signal-free intersections,'' \emph{Automatica}, vol.~93, pp. 244--256, 2018.

\bibitem{dresner2004multiagent}
K.~Dresner and P.~Stone, ``Multiagent traffic management: A reservation-based
  intersection control mechanism,'' in \emph{Autonomous Agents and Multiagent
  Systems, International Joint Conference on}, vol.~3.\hskip 1em plus 0.5em
  minus 0.4em\relax IEEE Computer Society, 2004, pp. 530--537.

\bibitem{huang2012assessing}
S.~Huang, A.~W. Sadek, and Y.~Zhao, ``Assessing the mobility and environmental
  benefits of reservation-based intelligent intersections using an integrated
  simulator,'' \emph{IEEE Transactions on Intelligent Transportation Systems},
  vol.~13, no.~3, pp. 1201--1214, 2012.

\bibitem{treiber2000}
\BIBentryALTinterwordspacing
M.~Treiber, A.~Hennecke, and D.~Helbing, ``Congested traffic states in
  empirical observations and microscopic simulations,'' \emph{Phys. Rev. E},
  vol.~62, pp. 1805--1824, Aug 2000. [Online]. Available:
  \url{https://link.aps.org/doi/10.1103/PhysRevE.62.1805}
\BIBentrySTDinterwordspacing

\bibitem{Katsikopoulos2003}
K.~V. {Katsikopoulos} and S.~E. {Engelbrecht}, ``Markov decision processes with
  delays and asynchronous cost collection,'' \emph{IEEE Transactions on
  Automatic Control}, vol.~48, no.~4, pp. 568--574, 2003.

\bibitem{bertsekas1995dynamic}
D.~P. Bertsekas, D.~P. Bertsekas, D.~P. Bertsekas, and D.~P. Bertsekas,
  \emph{Dynamic programming and optimal control}.\hskip 1em plus 0.5em minus
  0.4em\relax Athena scientific Belmont, MA, 1995, vol.~1, no.~2.

\bibitem{watkins1992q}
C.~J. Watkins and P.~Dayan, ``Q-learning,'' \emph{Machine learning}, vol.~8,
  no. 3-4, pp. 279--292, 1992.

\end{thebibliography}
\end{document}